\documentclass[aps,prb,preprint,superscriptaddress,floatfix]{revtex4}

\usepackage{graphicx,color}

\newcommand{\rev}[1]{\textcolor{black}{#1}}

\graphicspath{{figs/}}
\begin{document}
\title{Topologically protected interface phonons in two-dimensional nanomaterials: hexagonal boron nitride and silicon carbide}
\author{Jin-Wu Jiang}
    \altaffiliation{Corresponding author: jwjiang5918@hotmail.com}
    \affiliation{Shanghai Institute of Applied Mathematics and Mechanics, Shanghai Key Laboratory of Mechanics in Energy Engineering, Shanghai University, Shanghai 200072, People's Republic of China}
\author{Bing-Shen Wang}
    \affiliation{State Key Laboratory of Semiconductor Superlattice and Microstructure and Institute of Semiconductor, Chinese Academy of Sciences, Beijing 100083, China}
\author{Harold S. Park}
    \altaffiliation{Corresponding author: parkhs@bu.edu}
    \affiliation{Department of Mechanical Engineering, Boston University, Boston, Massachusetts 02215, USA}

\date{\today}
\begin{abstract}

We perform both lattice dynamics analysis and molecular dynamics simulations to demonstrate the existence of topologically protected phonon modes in two-dimensional, monolayer hexagonal boron nitride and silicon carbide sheets. The topological phonon modes are found to be localized at an \rev{in-plane} interface that divides these systems into two regions of distinct valley Chern numbers. The dispersion of this topological phonon mode crosses over the frequency gap, which is opened through analogy with the quantum valley Hall effect by breaking inversion symmetry of the primitive unit cells.  Consequently, vibrational energy with frequency within this gap is topologically protected, resulting in wave propagation that exhibits minimal backscattering, is robust with regards to structural defects such as sharp corners, and exhibits excellent temporal stability.  Our findings open up the possibility of actuating and detecting topological phonons in two-dimensional nanomaterials.

\end{abstract}

\maketitle
\pagebreak

\section{Introduction}

Over the past decade, there has been significant interest in a new state of matter, called topological insulators (TIs), whose behavior depends on its topology, rather than its geometry.  The distinguishing feature of TIs is that conducting edge or surface states are topologically protected,\cite{KaneCL2005prl,HasanMZ2010rmp,WangJ2017nm} where the TI is an insulator in the bulk while simultaneously allowing wave propagation along its boundary.  While the study of TIs originated in quantum electronic systems~\cite{HasanMZ2010rmp,mooreNATURE2010,qiPT2010,bernevigSCIENCE2006,kanePRL2005a}, the concept of topological protection has been extended to analyze other physical properties using classical principles. For instance, topologically protected edge states in photonic crystals can be discussed in analogy with the quantum Hall edge states,\cite{HaldaneFDM2008prl,LuL2014npho,ChenXD2017prb} and the interaction between photons and phonons can produce a Chern insulator of different topological phases.\cite{PeanoV2015prx} 

Recently, researchers have found that the topological nature of mechanical systems can also be investigated using analogs from electronic TIs. Efforts to control and guide phononic wave energy has led to various studies on phononic TIs based on the quantum hall effect~\cite{swinteckJAP2015,nassarJMPS2017,prodanPRL2009,nashPNAS2015,wangPRL2015,kariyadoSR2015,yangPRL2015,chenPRA2016,khanikaevNC2015}, quantum spin hall effect~\cite{mousaviNC2015,susstrunkPNAS2016,susstrunkSCIENCE2015,palJAP2016,palARXIV2017,yuARXIV2017,prodanNC2017,husseinAMR2014,heNP2016,cummerNRM2016,palNJP2017,xiaoNP2015}, and quantum valley hall effect~\cite{renRPP2016,palNJP2017,liuARXIV2017,wuARXIV2017}.  Other researchers have focused on investigating topological aspects of phonon modes.\cite{KosevichAM2004ltp,ZhangL2010prl,SusstrunkR2016pnas,LiuY2017prb}  Topologically protected elastic waves have also been observed in metamaterials or designed lattice models.\cite{mousaviNC2015,YangZ2015prl,khanikaevNC2015,PalRK2016jap,SussmanDM2016sm,JiWC2017cpl}  The topological concept was used to analyze zero-energy edge modes (`floppy modes') in lattice models,\cite{SunK2012pnas,KaneCL2014nphy,RocklinDZ2017nc} dynamic edge modes in biological or mechanical systems,\cite{ProdanE2009prl,BergN2011pre,ProdanE2017nc} topological modes localized at dislocations in mechanical metamaterials,\cite{PauloseJ2015nphy,LubenskyTC2015rpp} and the selective buckling via the states of self-stress analogous to topological quantum states in the two- and three-dimensional metamaterials built out of stacked kagome lattice models.\cite{PauloseJ2015pnas}  Mechanical lattice models can also display topologically protected zero-energy phonon modes, which are analogs of massless fermion states of topological Weyl or nodal semimetals.\cite{RocklinDZ2016prl,StenullO2016prl,PoHC2016prb}

However, much of the above research on topological phonons and phononic TIs has focused on either discrete lattice models or macroscale structures, while very few works have considered topological phonons in nanomaterials.  This is in contrast to various works that have investigated topologically protected electronic and spin conduction in nanomaterials, such as the valley-contrasting topological transport in graphene,\cite{XiaoD2007prl} the spontaneous quantum Hall states in chirally stacked few-layer graphene systems,\cite{ZhangF2011prl} and the electronic edge modes and the topological transitions in bilayer graphene.\cite{ZhangF2013pnas} 

In this paper, we demonstrate topologically protected phonon transport in two-dimensional nanomaterials, monolayer hexagonal boron nitride (h-BN) and silicon carbide (SiC).  We study, using lattice dynamics analysis and classical molecular dynamics (MD) simulations, a specific set of phonon modes in h-BN and SiC, i.e., the topological phonon modes, which can be localized at an \rev{in-plane} interface connecting topologically trivial and non-trivial structures.  By utilizing concepts from the quantum valley Hall effect, we demonstrate the formation of a frequency gap [1123, 1278]~{cm$^{-1}$} in the phonon spectrum of the h-BN and [126.7, 348.6]~{cm$^{-1}$} in the phonon spectrum of SiC, due to the broken inversion symmetry of their primitive unit cells. We observe a topological phonon branch at the interface that divides the h-BN and SiC into regions of trivial and non-trivial topologies. This topological phonon branch crosses over the frequency gap [1123, 1278]~{cm$^{-1}$} for h-BN or [126.7, 348.6]~{cm$^{-1}$} for SiC in the phonon spectrum, so vibrational energy within this frequency gap can only be transported by the topological phonon modes. This energy transfer is topologically protected, and we demonstrate through various examples that it is both spatially and temporally robust.

The present paper is organized as follows. In Sec.~II, we present some details for the phonon analysis and MD simulation. Section III is devoted to the phonon analysis for h-BN and SiC, while topological discussions are presented in Sec.~IV. MD simulation results are presented in Sec.~V. The paper ends with a brief summary in Sec.~VI.

\section{Computational details}

We investigate the phonon dispersion of three different 2D materials:  graphene, h-BN, and SiC.  Graphene is considered to illustrate the phonon dispersion when inversion symmetry exists in a 2D material with a hexagonal lattice structure. The carbon-carbon interactions in graphene were described by the Brenner potential.\cite{brennerJPCM2002} The interatomic interactions in h-BN are described by the Tersoff potential,\cite{LindsayL2011prb} while the interatomic interactions in SiC are also described by the Tersoff potential.\cite{TersoffJ5}

\rev{The phonon dispersion and eigenvectors were computed using the package GULP,\cite{gulp} in which the dynamical matrix is calculated to compute the phonon dispersion,
\begin{eqnarray}
D_{s\alpha;s'\beta}\left(\vec{k}\right) & = & \frac{1}{\sqrt{m_{s}m_{s'}}}\sum_{l_{1}=1}^{N_{1}}\sum_{l_{2}=1}^{N_{2}}K_{00s\alpha;l_{1}l_{2}s'\beta}e^{i\vec{k}\cdot\vec{R}_{l_{1}l_{2}}},\nonumber\\
\end{eqnarray}
where $\vec{R}_{l_{1}l_{2}}$ is the lattice vector, and $m_s$ is the mass for the atom s in the unit cell. Usually, the summation over lattice sites $(l_1, l_2)$ can be truncated to the summation over neighboring atoms in case of short-range interactions. The force constant matrix is $K_{l_{1} l_{2} s\alpha;l_{1}' l_{2}' s'\beta}=\frac{\partial^{2} V}{\partial u_{l_{1} l_{2} s}^{\alpha} \partial u_{l_{1}' l_{2}' s'}^{\beta}}$ with $V$ as the atomic interaction. For the phonon mode $\tau$ at the wave vector $\vec{k}$, the phonon dispersion $\omega^{(\tau)2}(\vec{k})$ and eigenvectors $\vec{\xi}^{(\tau')}$ were computed from the eigenvalue solution of the dynamical matrix,
\begin{eqnarray}
\sum_{s'\beta}D_{s\alpha;s'\beta}\left(\vec{k}\right)\xi_{\beta}^{(\tau')}(\vec{k}|00s') & = & \omega^{(\tau)2}(\vec{k})\xi_{\alpha}^{(\tau')}(\vec{k}|00s).
\end{eqnarray}
}

For MD simulations, the standard Newton equations of motion are integrated in time using the velocity Verlet algorithm with a time step of 1~{fs}. Simulations are performed using the publicly available simulation code LAMMPS~\cite{PlimptonSJ}, while the OVITO package is used for visualization~\cite{ovito}.

\begin{figure*}[htpb]
  \begin{center}
    \scalebox{1.0}[1.0]{\includegraphics[width=\textwidth]{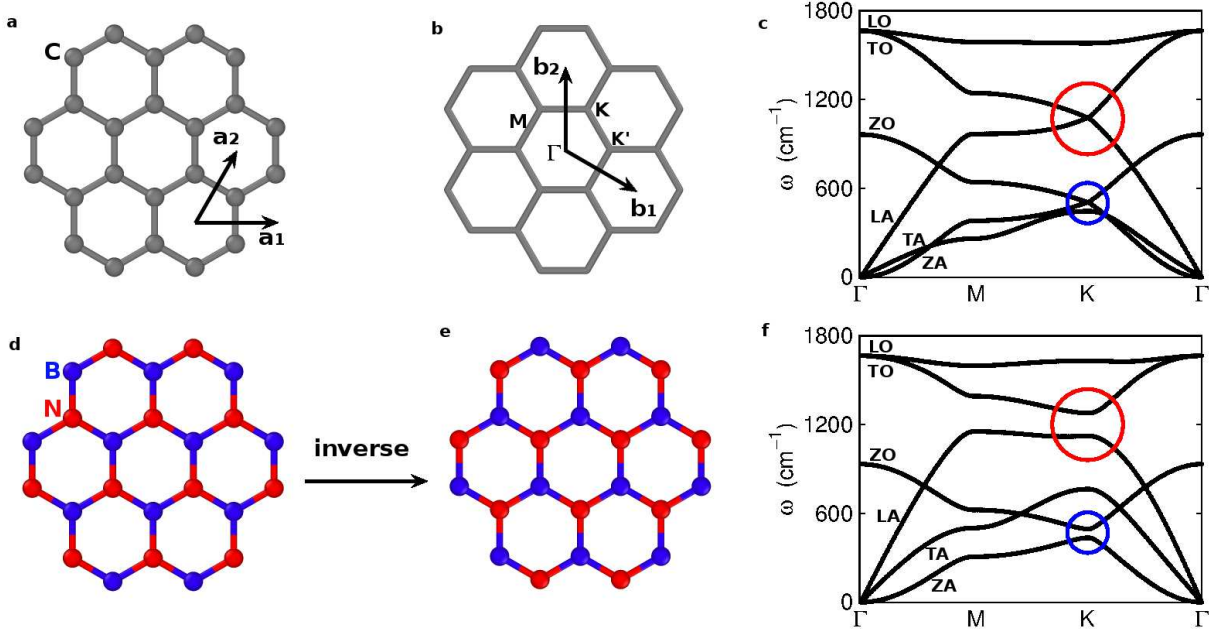}}
  \end{center}
  \caption{(Color online) Phonon dispersion for monolayer graphene and monolayer h-BN. (a) Hexagonal lattice structure of graphene of D$_{\rm 6h}$ symmetry, including the inversion center. (b) The hexagonal reciprocal lattice. $\Gamma$, M, and K are high symmetry points in the first Brillouin zone. (c) Phonon dispersion for graphene. The red and blue circles highlight two instances of Dirac-like dispersion at the K point. (d) and (e): Lattice structure of h-BN of D$_{\rm 3h}$ symmetry. The inversion symmetry, which switches the B and N atoms, is broken. (f) Phonon dispersion for h-BN. The red and blue circles highlight the gap opening at the Dirac-like point.}
  \label{fig_phonon_bulk}
\end{figure*}

\begin{figure}[htpb]
  \begin{center}
    \scalebox{1.0}[1.0]{\includegraphics[width=8cm]{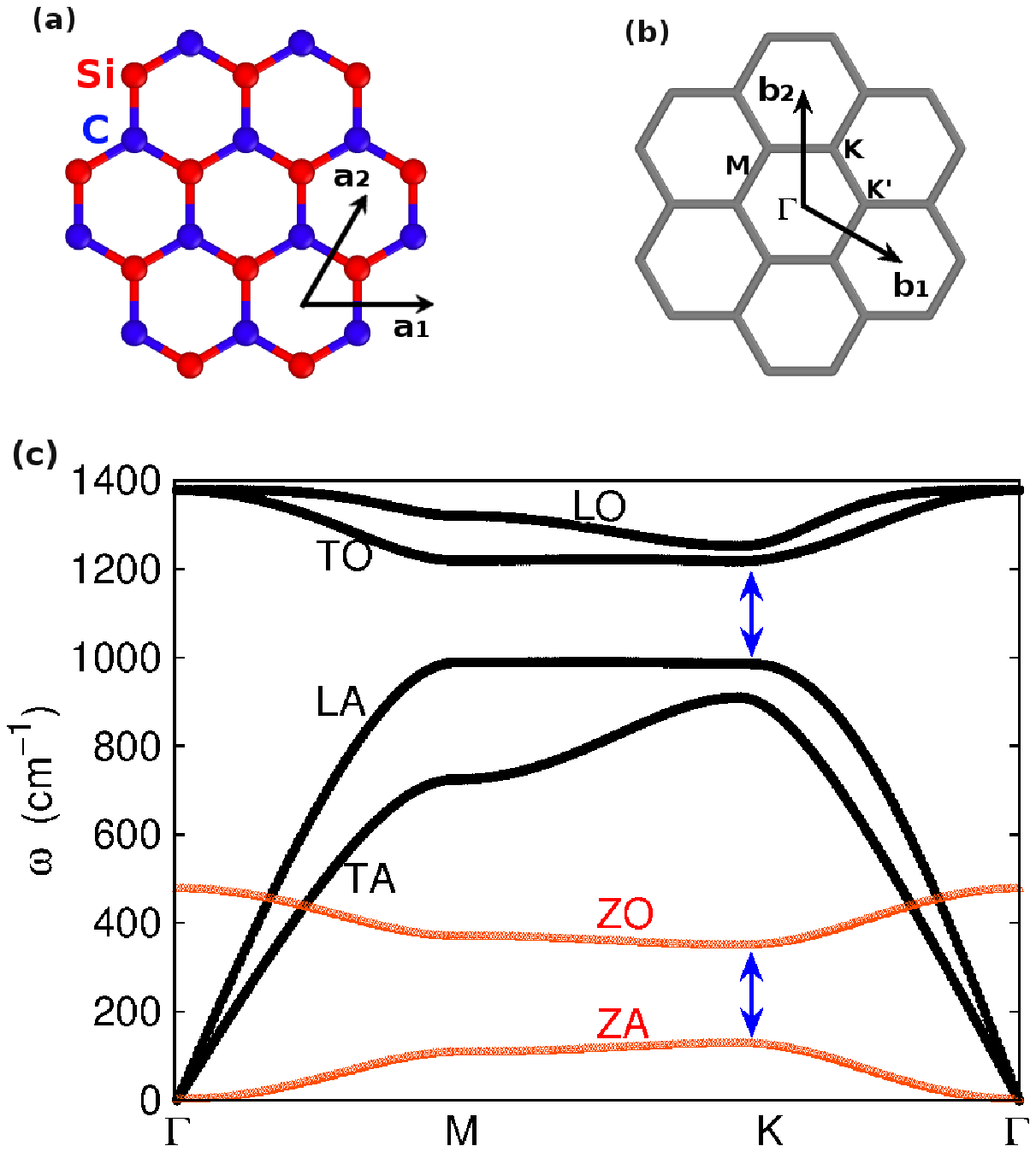}}
  \end{center}
  \caption{(Color online) Phonon dispersion for monolayer SiC. (a) Hexagonal lattice structure of SiC of D$_{\rm 3h}$ symmetry. The inversion symmetry, which switches the Si and C atoms, is broken. (b) The hexagonal reciprocal lattice. $\Gamma$, M, and K are high symmetry points in the first Brillouin zone. (c) Phonon dispersion for SiC. The arrows highlight the gap opening at the K point.}
  \label{fig_phonon_bulk_sic}
\end{figure}

\section{Phonon dispersion analysis}

\subsection{Phonon dispersion for h-BN}
Graphene has a honeycomb lattice structure of D$_{\rm 6h}$ symmetry as shown in Fig.~\ref{fig_phonon_bulk}~(a). The primitive unit cell is denoted by two basis vectors $\vec{a}_1=a\hat{e}_x$ and $\vec{a}_2=a(\frac{1}{2}\hat{e}_x + \frac{\sqrt{3}}{2}\hat{e}_y)$, with $a=1.42$~{\AA} as the lattice constant. The x-axis is in the horizontal direction, while the y-axis is in the vertical direction. The two (carbon) atoms in the primitive unit cell are the same, so inversion symmetry exists in this structure. Fig.~\ref{fig_phonon_bulk}~(b) shows the reciprocal space, which is also a hexagonal lattice structure, with two basis vectors $\vec{b}_1=b(\frac{\sqrt{3}}{2}\hat{e}_x - \frac{1}{2}\hat{e}_y)$ and $\vec{b}_2=b\hat{e}_y$ and $b=\frac{4\pi}{\sqrt{3}a}$.

Fig.~\ref{fig_phonon_bulk}~(d) shows the hexagonal lattice structure of h-BN. Different from graphene, the two atoms (B and N) in the primitive unit cell are different, so the inversion symmetry is broken for h-BN. The inverted structure for h-BN shown in Fig.~\ref{fig_phonon_bulk}~(e) is different from its original structure in Fig.~\ref{fig_phonon_bulk}~(d).

The phonon dispersion for monolayer graphene is shown in Fig.~\ref{fig_phonon_bulk}~(c) along the high symmetry $\Gamma$MK$\Gamma$ lines in the first Brillouin zone. \rev{There are six phonon branches corresponding to the two atoms in the primitive unit cell, i.e., the z-directional acoustic (ZA), the transverse acoustic (TA), the longitudinal acoustic (LA), the z-directional optical (ZO), the transverse optical (TO), and the longitudinal optical (LO) branches.} There are two Dirac-like dispersions (depicted by red and blue circles) at the K point, which are both gapless. In other words, the phonon modes at these two frequencies are degenerate, resulting from the inversion symmetry of the two carbon atoms in the primitive unit cell for graphene.

Fig.~\ref{fig_phonon_bulk}~(f) shows the phonon dispersion for monolayer h-BN. Compared with the phonon dispersion of graphene, a distinct feature is the opening of the frequency gaps for the two Dirac-like dispersions, as a result of the broken inversion symmetry for the primitive unit cell of h-BN. The higher frequency gap [1123, 1278]~{cm$^{-1}$} is of particular importance, because there is no other phonon branch falling within this frequency gap. Hence, vibrational energy with frequency in this gap cannot be transported in h-BN. However, if an interface phonon branch crossing over the frequency gap can be generated, then these interface modes will be topologically protected against different backscattering mechanisms. 

\subsection{Phonon dispersion for SiC}

SiC has a honeycomb lattice structure of D$_{\rm 3h}$ symmetry as shown in Fig.~\ref{fig_phonon_bulk_sic}~(a), which is similar as the structure of h-BN. The lattice constant is $a=3.121$~{\AA} from the Tersoff potential. Fig.~\ref{fig_phonon_bulk_sic}~(c) shows the phonon dispersion for monolayer SiC. Similar to h-BN, a distinct feature in the phonon dispersion is the opening of the frequency gaps for the two Dirac-like dispersions at the K point, as a result of the broken inversion symmetry for the primitive unit cell of SiC. The higher frequency gap locates at [986.2, 1217.8]~{cm$^{-1}$}, which corresponds to the in-plane vibrations. The lower frequency gap is at [126.7, 348.6]~{cm$^{-1}$}, which corresponds to out-of-plane vibrations. The value of the frequency gap is 231.6~{cm$^{-1}$} and 221.9~{cm$^{-1}$} for these two frequency gaps, both of which are larger than the frequency gap in the h-BN.  This is because the frequency gap is proportional to the mass difference of the two atoms in the primitive unit cell, and the mass difference between Si and C atoms in SiC is much larger than the B and N atoms in h-BN.

\begin{figure}[htpb]
  \begin{center}
    \scalebox{1}[1]{\includegraphics[width=8cm]{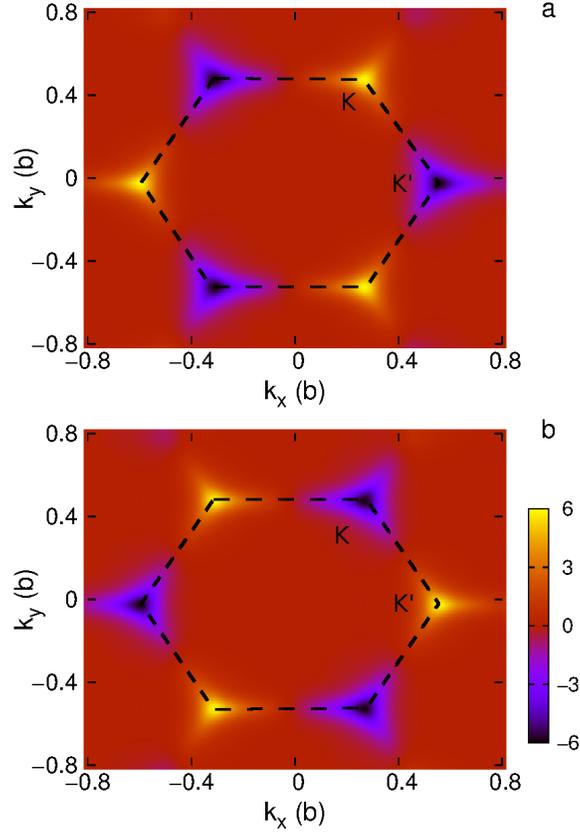}}
  \end{center}
  \caption{(Color online) Berry curvature for the lower boundary phonons of the frequency gap (i.e., the fourth branch) for (a) h-BN and (b) h-NB. The Berry curvature is localized at the K and K$'$ points, and is opposite in h-BN and h-NB.}
  \label{fig_berry}
\end{figure}

\begin{figure*}[htpb]
  \begin{center}
    \scalebox{1.0}[1.0]{\includegraphics[width=\textwidth]{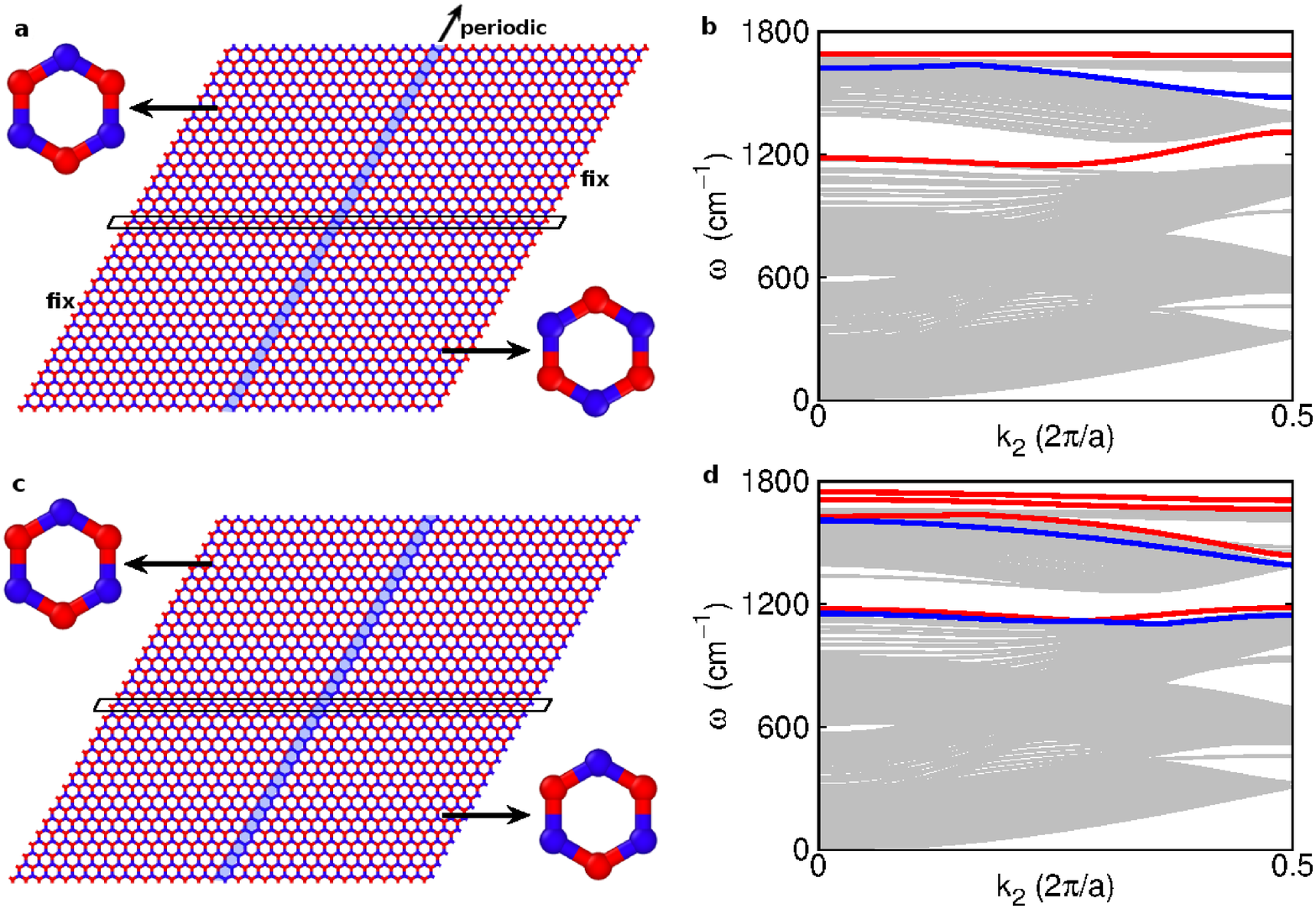}}
  \end{center}
  \caption{(Color online) Phonon dispersion for the topological and trivial interfaces (blue areas) in the monolayer h-BN ribbon. (a) The configuration of a topological interface in the middle of the structure along the $\vec{a}_2$ direction. The left and right regions that intersect at the interface are of different topologies. (b) Phonon dispersion for the topological interface. Edge (interface) branches are depicted by blue (red) thick lines. Note the topological interface phonon branch crossing over the frequency gap [1123, 1278]~{cm$^{-1}$}. (c) The structure of a topologically trivial interface along the $\vec{a}_2$ direction. The left and right regions that intersect at the interface are of the same topology. (d) Phonon dispersion for the trivial interface. There is no phonon branch crossing over the frequency gap [1123, 1278]~{cm$^{-1}$}.}
  \label{fig_phonon_interface}
\end{figure*}

\section{Topological analysis}

\subsection{Topological analysis for h-BN}

We further show in Fig.~\ref{fig_berry} the Berry curvature of the phonon modes at the lower boundary of the frequency gap. The Berry curvature for the phonon mode indexed by $\tau$ at the wave vector $\vec{k}$ is calculated by\cite{PalRK2017njp}
\begin{eqnarray}
B^{\tau}\left(\vec{k}\right) = -2Im\sum_{\tau'\not=\tau}\frac{\langle\tau|\frac{\partial D}{\partial k_{x}}|\tau'\rangle\langle\tau'|\frac{\partial D}{\partial k_{y}}|\tau\rangle}{\left(\omega_{\tau}^{2}-\omega_{\tau'}^{2}\right)^{2}},
\label{eq_berry}
\end{eqnarray}
where $D$ is the dynamical matrix. $\omega_{\tau}$ and $|\tau\rangle$ are the frequency and the polarization vector of the phonon mode $\tau$, respectively. Fig.~\ref{fig_berry}~(a) is the Berry curvature for the lower boundary phonons of the frequency gap in the h-BN sheet, which is localized at the K and K$'$ points in the first Brillouin zone. Fig.~\ref{fig_berry}~(b) is the Berry curvature for the lower boundary phonons of the frequency gap in the h-NB (i.e., B and N atoms are exchanged as compared with h-BN), which has opposite sign as compared with h-BN in Fig.~\ref{fig_berry}~(a). The valley Chern number is computed by integrating the Berry curvature over a small region near the K and K$'$ points as,
\begin{eqnarray}
C_{\nu}^{\tau} & = & \frac{1}{2\pi}\int_{\nu}B^{\tau}\left(\vec{k}\right)d\vec{k},
\label{eq_chern}
\end{eqnarray}
where $\nu=$ K, K$'$ is the valley index.  \rev{The Chern number is a topological invariant that characterizes the nature of the topological interface.  As discussed in previous works,\cite{palNJP2017} the Chern numbers should have opposite signs for the upper and lower bands of the lattices along the interface in order for bulk-boundary correspondance to guarantee the presence of topologically protected localized modes along the interface.}

For the h-BN lattice, the obtained Chern numbers are 0.28 and -0.28 for the K and K$'$ valleys, respectively. For the h-NB lattice, the Chern numbers are -0.28 and 0.28 for the K and K$'$ valleys, respectively. Consequently, topologically protected localized phonon modes exist at the interface of the h-BN and h-NB lattices with different valley Chern numbers, which is in analogy with the quantum valley hall effect.\cite{XiaoD2007prl,rycerzNP2007} It should be noted that the Chern numbers deviate from the ideal value of $\pm 0.5$, which corresponds to the highly localized Berry curvature around K and K$'$ points. It is because the extension of the Berry curvature in the reciprocal space as shown in Fig.~\ref{fig_berry} will lead to the reduction of the integral for the Chern number in Eq.~(\ref{eq_chern}).\cite{ZhuH2017arxiv}

We now demonstrate the generation of the topological phonon branch that crosses over the frequency gap [1123, 1278]~{cm$^{-1}$} for h-BN. Fig.~\ref{fig_phonon_interface}~(a) shows a monolayer h-BN sheet denoted by $n_1 \vec{a}_1 \times n_2 \vec{a}_2$. The structure shown in the figure has the size $32 \vec{a}_1 \times 32 \vec{a}_2$. The left and right ends are fixed for the phonon calculation, while periodic boundary conditions are applied along the $\vec{a}_2$ direction. The big unit cell enclosed by the black box is used for the  phonon calculation. There is an interface along the $\vec{a}_2$ direction in the middle of the structure, which divides the structure into the left and right regions with different topology (see left top and right bottom insets). This type of interface will be referred to as the topological interface. \rev{The B-B or N-N bonds at the interface in the h-BN are homoelemental bonds, which were predicted to exist in some defected h-BN sheets. The stability of the B-B or N-N interfaces has been investigated by previous first-principles calculations.\cite{LiuY2012acsn}} Fig.~\ref{fig_phonon_interface}~(b) displays the phonon dispersion for the topological interface shown in Fig.~\ref{fig_phonon_interface}~(a). There are 192 branches corresponding to the 64 atoms in the big unit cell. The edge modes localized nearby the two fixed ends are denoted by blue lines. The red lines depict the phonon modes localized at the interface. There is a particular interface phonon branch crossing over the frequency gap [1123, 1278]~{cm$^{-1}$}.

The h-BN lattice structure shown in Fig.~\ref{fig_phonon_interface}~(c) has the same size of that in Fig.~\ref{fig_phonon_interface}~(a). The width of the interface is just one column of atoms thicker than the topological interface shown in Fig.~\ref{fig_phonon_interface}~(a). As a result of this difference, the left and right areas around the interface have the same topology (see left top and right bottom insets). We will thus refer to this interface as the trivial interface. There is no phonon branch crossing over the frequency gap [1123, 1278]~{cm$^{-1}$} in the phonon dispersion for this trivial interface shown in Fig.~\ref{fig_phonon_interface}~(d).

\begin{figure*}[htpb]
  \begin{center}
    \scalebox{1}[1]{\includegraphics[width=\textwidth]{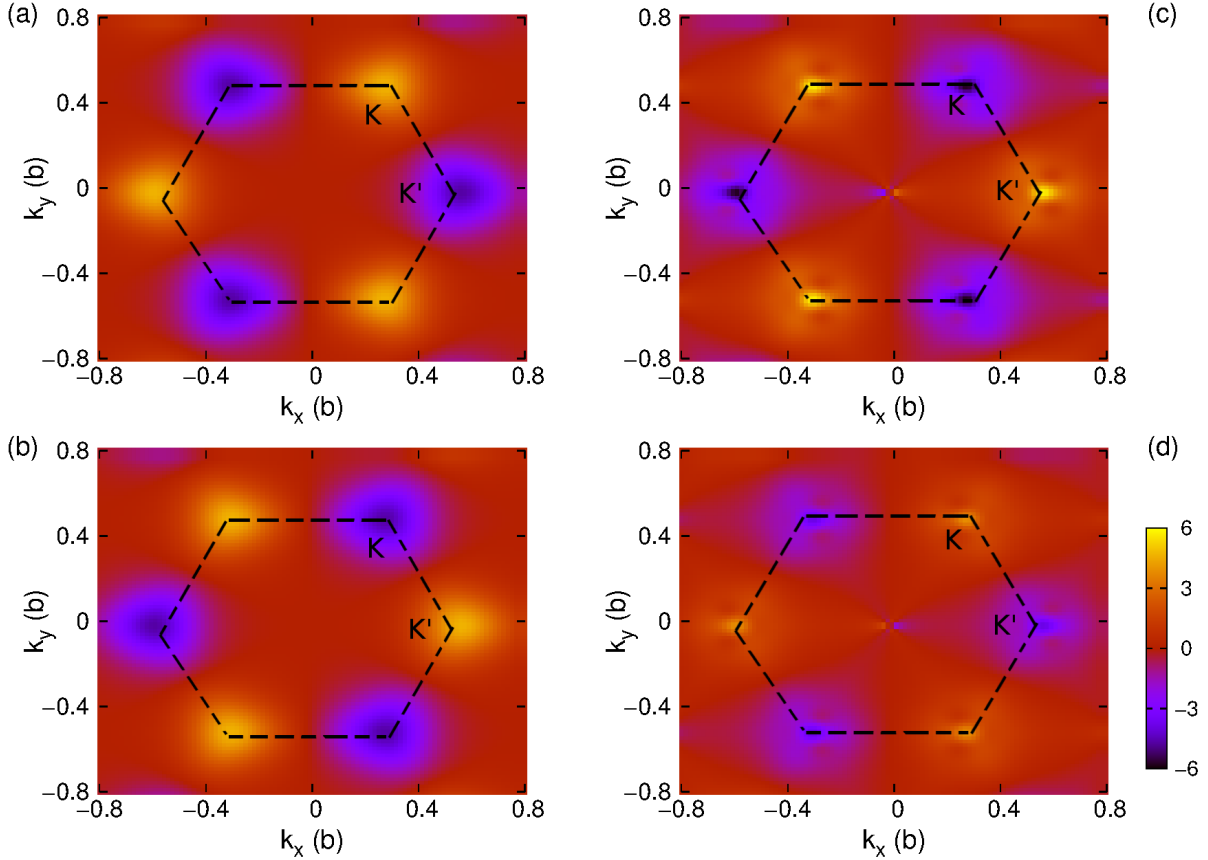}}
  \end{center}
  \caption{(Color online) Berry curvature for the ZA branch of (a) SiC and (b) CSi, and TO branch of (c) SiC and (d) CSi. The Berry curvature is localized at the K and K$'$ points, and is opposite in SiC and CSi.}
  \label{fig_berry_sic}
\end{figure*}

\begin{figure}[htpb]
  \begin{center}
    \scalebox{1.0}[1.0]{\includegraphics[width=8cm]{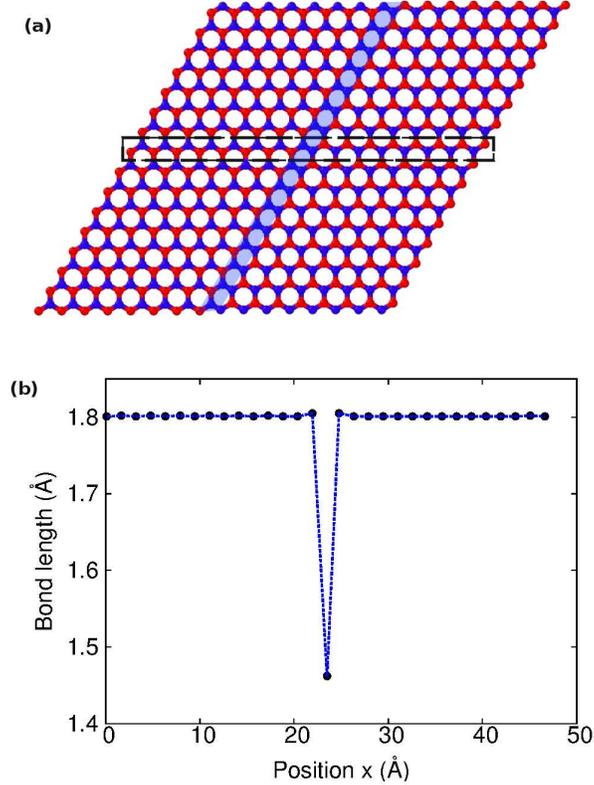}}
  \end{center}
  \caption{(Color online) Structure of the C-C interface. (a) The C-C interface (blue area) is in the middle of the crystal and along the $\vec{a}_2$ direction. The rectangular box denotes the translational cell along the $\vec{a}_2$ direction. (b) The distribution of the bond length along the x-direction (horizontal).}
  \label{fig_bond_length_sic}
\end{figure}

\begin{figure*}[htpb]
  \begin{center}
    \scalebox{1.0}[1.0]{\includegraphics[width=\textwidth]{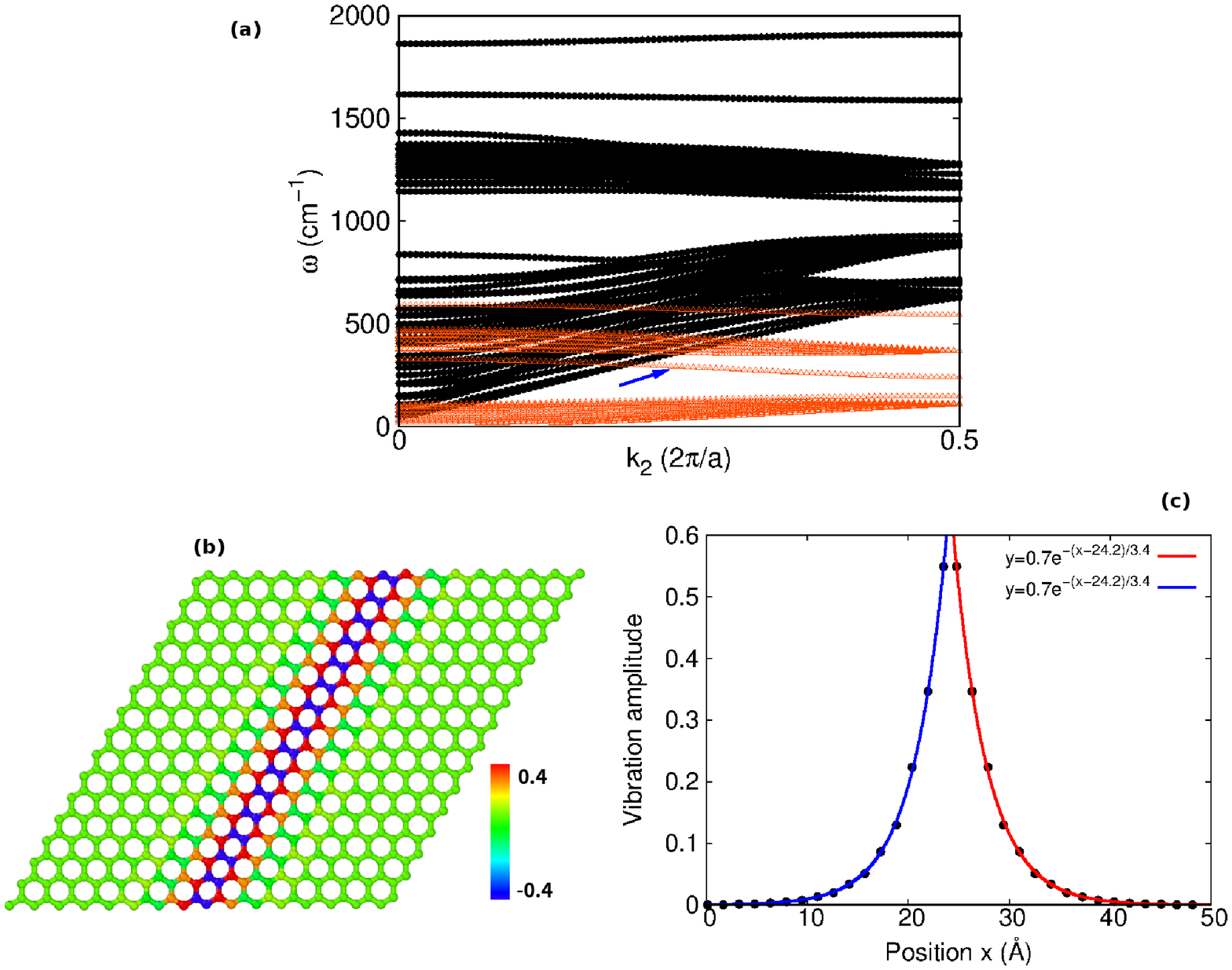}}
  \end{center}
  \caption{(Color online) Phonon dispersion for the interface in the monolayer SiC ribbon. (a) Phonon dispersion for the SiC ribbon with C-C interface. The xy (z) directional phonons are denoted by black (red) lines. The interface mode (depicted by blue arrow) crosses over the frequency gap of the z-directional phonons. (b) The eigenvector of the interface mode at $k_2=0$. The color bar represents the value of the eigenvector. (c) An exponential decay (from the interface) for the amplitude of the eigenvector in (b).}
  \label{fig_phonon_interface_sic}
\end{figure*}

\subsection{Topological analysis for SiC}

We now discuss the topological properties of the phonon modes in SiC. Fig.~\ref{fig_berry_sic} shows the Berry curvature of the ZA and TO branches in SiC. Fig.~\ref{fig_berry_sic}~(a) is the Berry curvature for the lower boundary phonons of the low-frequency gap in the SiC sheet, which is localized at the K and K$'$ points in the first Brillouin zone. Fig.~\ref{fig_berry_sic}~(b) is the Berry curvature for the lower boundary phonons of the low-frequency gap in the CSi (i.e., Si and C atoms are exchanged as compared with SiC), which has opposite sign as compared with SiC in Fig.~\ref{fig_berry_sic}~(a). For the SiC lattice, the obtained Chern numbers from Eq.~(\ref{eq_chern}) are 0.21 and -0.21 for the K and K$'$ valleys, respectively. For the CSi lattice, the Chern numbers are -0.21 and 0.21 for the K and K$'$ valleys, respectively. These values are smaller than the Chern numbers ($\pm 0.28$) for h-BN, because the interface in CSi is more distorted than that of h-BN. In other words, the frequency gap opened at the K point for SiC is larger than that of h-BN. Topologically protected localized phonon modes shall exist at the interface of the SiC and CSi lattices with different valley Chern numbers, which is in analogy with the quantum valley hall effect.\cite{XiaoD2007prl,rycerzNP2007} Figs.~\ref{fig_berry_sic}~(c) and (d) show similar phenomenon for the top boundary phonon of the high-frequency gap.

We examine the configuration details for the C-C interface in Fig.~\ref{fig_bond_length_sic}. The C-C interface is in the middle of the lattice and along the $\vec{a}_2$ direction in Fig.~\ref{fig_bond_length_sic}~(a). The structure is denoted by $n_1 \vec{a}_1 \times n_2 \vec{a}_2$. The structure shown in the figure has the size $16 \vec{a}_1 \times 16 \vec{a}_2$. The carbon atom is much smaller than the silicon atom, so the C-C bond length is considerably shorter than the Si-C bond. Fig.~\ref{fig_bond_length_sic}~(b) shows the distribution of the bond length along the x-direction, where the C-C bond length is about 1.462~{\AA}. Furthermore, due to the large difference between the C-C and Si-C bonds, there are obvious distortions for the Si-C bond length close to the interface. \rev{First-principles calculations have predicted a stable planar structure for the SiC in the two-dimensional configuration,\cite{SunL2008jchemp} and the stability of the two-dimensional Si$_{x}$C$_{1-x}$ monolayers (including various interfaces) with $0\le x\le 1$ has been investigated by first-principles calculations.\cite{ShiZ2015acsn}}

We now demonstrate the generation of the topological phonon branch that crosses over the frequency gaps for SiC. Fig.~\ref{fig_phonon_interface_sic}~(a) displays the phonon dispersion for the SiC ribbon with C-C interface shown in Fig.~\ref{fig_bond_length_sic}~(a). The left and right ends are fixed for the phonon calculation, while periodic boundary conditions are applied along the $\vec{a}_2$ direction. The big unit cell enclosed by the black box is used for the  phonon calculation. There are 96 branches corresponding to the 32 atoms in the big unit cell. There is a particular interface phonon branch (indicated by the blue arrow) crossing over the lower-frequency gap [126.7, 348.6]~{cm$^{-1}$}. There is no interface phonon crossing over the higher-frequency gap [986.2, 1217.8]~{cm$^{-1}$}, which is probably because of the large difference between the strength of the C-C and the Si-C bonds.  As discussed above, the C-C bond is much shorter than the Si-C bond, which indicates that the C-C bond is much stronger than the Si-C bond. The vibration of the interface phonon mainly involves the vibration of the C-C bond, so its frequency is much higher than the vibration of the Si-C bond; i.e., the interface phonon branch with in-plane vibrations is out of the high-frequency range of the SiC. However, the C-C and Si-C bonds have similar chemical properties in the out-of-plane direction, due to the constraint of the honeycomb lattice structure. As a result, the interface phonon with out-of-plane vibrations can cross over the lower-frequency gap.

Fig.~\ref{fig_phonon_interface_sic}~(b) illustrates the eigenvector for the interface phonon with frequency $\omega=325.8$~{cm$^{-1}$} at $k_2=0$ point. Only atoms at the interface are involved in this eigenvector, so the interface phonon is a kind of localized mode. Fig.~\ref{fig_phonon_interface_sic}~(c) clearly displays the exponential decay of the vibrational component away from the interface, which is a characteristic property of the localized mode. It should be noted that the localization here is with respective to the x-direction (or $\vec{a}_1$ direction). The interface phonon has nonzero group velocity as can be seen from Fig.~\ref{fig_phonon_interface_sic}~(a), so the interface phonon can travel along the interface ($\vec{a}_2$ direction).

\begin{figure}[htpb]
  \begin{center}
    \scalebox{1.0}[1.0]{\includegraphics[width=8cm]{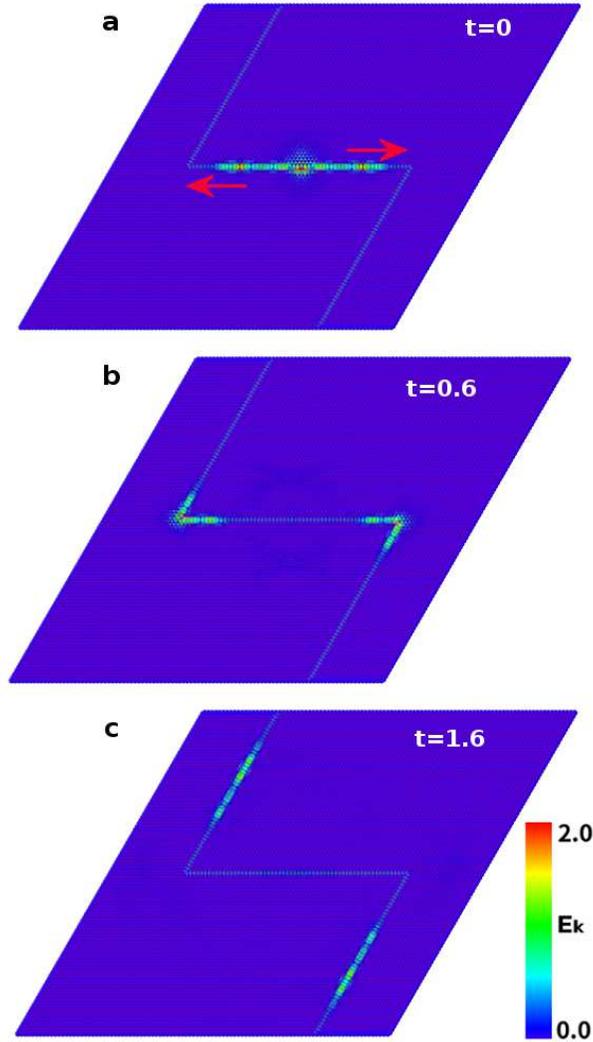}}
  \end{center}
  \caption{(Color online) Kinetic energy transfer along the zigzag shaped topological interface in monolayer h-BN at time (a) t=0~ps, (b) t=0.6~ps, and (c) t=1.6~ps. Note that the energy can be transported along the topological interface with minimal loss, even at the sharp corners of the zigzag. The color bar shows the kinetic energy (in meV) of each atom.}
  \label{fig_md_zigzag}
\end{figure}

\begin{figure}[htpb]
  \begin{center}
    \scalebox{1.0}[1.0]{\includegraphics[width=8cm]{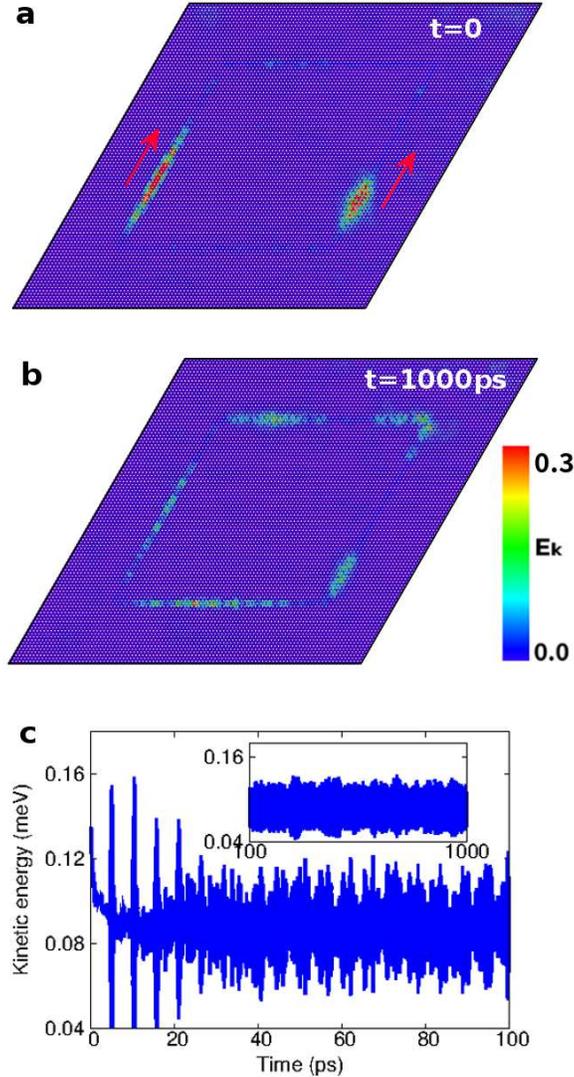}}
  \end{center}
  \caption{(Color online) Demonstration of long-time localization of the energy in a closed parallelogram topological interface in monolayer h-BN. (a) Two energy pulses are created at $t=0$. (b) A large portion of the kinetic energy is still localized at the interface after a long time (1000~ps). The color bar represents the kinetic energy (in meV) for each atom. (c) The time history of the total kinetic energy localized at the interface.  The high peaks in the initial 20~ps correspond to the scattering of two moving energy pulses. Inset shows the time history at longer time scales.}
  \label{fig_md_parall}
\end{figure}

\begin{figure}[htpb]
  \begin{center}
    \scalebox{1.0}[1.0]{\includegraphics[width=8cm]{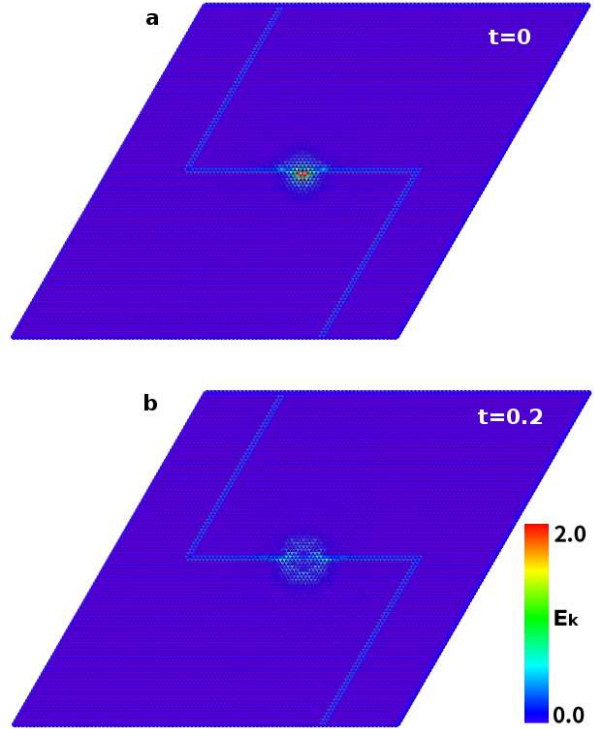}}
  \end{center}
  \caption{(Color online) Illustration of how kinetic energy is not transported along the zigzag shaped, topologically trivial interface in monolayer h-BN.}
  \label{fig_md_zigzag_trivial}
\end{figure}

\section{Molecular dynamics simulations}

In macroscopic systems, elastic waves are generated and detected along the topological interface. In contrast, for nanomaterials, the vibration energy transport can be simulated by molecular dynamics (MD) simulations. We thus perform MD simulations in this section to theoretically illustrate some transport properties for the topological phonons. In practice, there are several available approaches to actuate the topological phonons at a specific frequency, like optical methods or neutron scattering approaches. The optical approach is suitable for phonons at the $\Gamma$ point with zero wave vector for the topological phonon branch. The topological phonons are localized interface modes with low symmetry, so they are both Raman and infra-red active modes. Hence, the topological phonons can be readily excited by the optical approaches. The neutron scattering method is able to excite phonons of high frequency and at arbitrary wave vector in the Brillouin zone. The topological phonons may be investigated experimentally by these approaches.

\subsection{MD simulations for h-BN}

From the above, we have observed the topological phonon branch crossing over the frequency gap [1123, 1278]~{cm$^{-1}$} for the topological interface in monolayer h-BN.  Therefore, this topological phonon mode will be protected by the frequency gap due to the energy conservation law.  As a result, vibrational energy carried by the topological phonon mode will be highly stable and localized at the interface. To verify the stability and localization properties of the interface phonon mode, we performed MD simulations to study the energy transfer along the topologically trivial and non-trivial interfaces shown in Fig.~\ref{fig_phonon_interface} for monolayer h-BN.

We first simulate the energy transfer along the zigzag shaped topological interface shown in Fig.~\ref{fig_md_zigzag}, where the h-BN is divided into two areas of different topology. \rev{As a result, the zigzag interface corresponds to the topological interface shown in Fig.~\ref{fig_phonon_interface}~(a).} The atom in the center of the interface is driven to oscillate along the interface direction at a given frequency $\omega=1250$~{cm$^{-1}$} for 30 cycles, after which the system is allowed to evolve within the NVE (i.e., the particles number N, the volume V, and the energy E of the system are constant) ensemble. In these MD simulations, the topological phonons can be directly actuated at a chosen frequency. The resultant distribution of the kinetic energy in the h-BN is shown in Fig.~\ref{fig_md_zigzag}~(a). Two energy pulses, traveling in opposite directions, are generated. Figs.~\ref{fig_md_zigzag}~(b) and (c) illustrate a stable transfer of the kinetic energy along the interface. There is almost no energy loss during the energy transfer, even at the sharp corners of the zigzag interface, and about 99.8\% of the kinetic energy propagates around the sharp corner.  

To further demonstrate the stability of the energy localized at the topological interface, we simulate the kinetic energy localized within a closed parallelogram shaped interface. In Fig.~\ref{fig_md_parall}~(a), two energy pulses are created, which are moving in opposite directions. Both energy pulses travel along the topological interface and these two energy pulses will scatter after about 4.7~ps. The scattering of these two energy pulses results in an obvious signal in the kinetic energy as shown in Fig.~\ref{fig_md_parall}~(c), which is the time history for the total kinetic energy for the whole closed interface area. Fig.~\ref{fig_md_parall}~(c) shows four obvious scatterings between the two moving energy pulses, after which the pulses are divided into many smaller energy pulses. There is almost no energy loss for the kinetic energy of the interface, even after a long simulation time of 1000~ps. Fig.~\ref{fig_md_parall}~(b) shows that the kinetic energy is still mainly localized at the interface after 1000~ps.

In contrast to the topological interface, Fig.~\ref{fig_md_zigzag_trivial} shows that there is almost no energy transfer along the zigzag shaped trivial interface, where the h-BN is divided into two areas of the same topology. The middle atom at the interface is driven to oscillate along the interface direction at a given frequency $\omega=1250$~{cm$^{-1}$} for 30 cycles, and Fig.~\ref{fig_md_zigzag_trivial}~(a) shows the resultant distribution of the kinetic energy, which clearly cannot be transferred along the trivial interface. The kinetic energy localized around the driving region will be dissipated into the regions surrounding the interface because this frequency is not topologically protected as shown in Fig.~\ref{fig_md_zigzag_trivial}~(b).

\subsection{MD simulations for SiC}

\begin{figure}[htpb]
  \begin{center}
    \scalebox{1.9}[1.9]{\includegraphics[width=8cm]{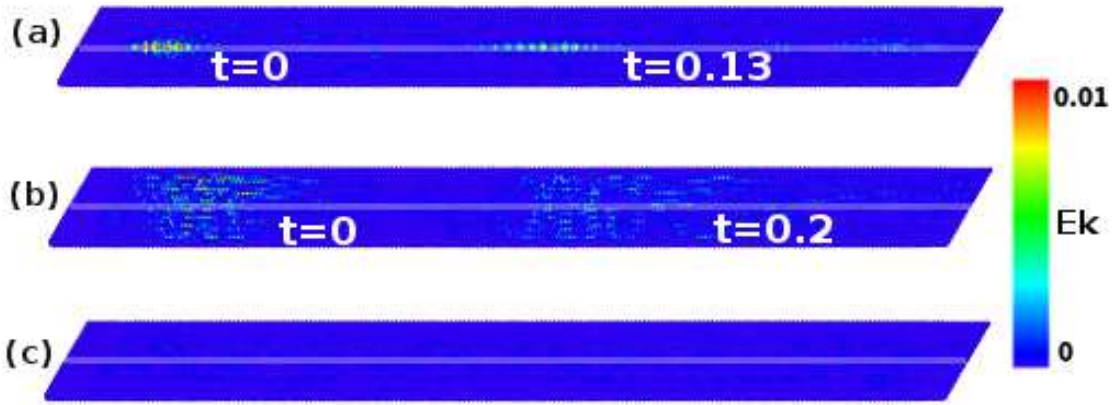}}
  \end{center}
  \caption{(Color online) Snapshots from MD for the transfer of the vibrational energy. The energy is injected into the SiC by vibrating one atom at the interface at a frequency of (a) 300~{cm$^{-1}$}, (b) 100~{cm$^{-1}$}, and (c) 700~{cm$^{-1}$}. The color bar represents the kinetic energy (in meV) for each atom. The time $t$ is in ps.}
  \label{fig_md_sic}
\end{figure}

From the above, we have observed the topological phonon branch crossing over the frequency gap [126.7, 348.6]~{cm$^{-1}$} for the C-C interface in monolayer SiC. As a result, vibrational energy carried by the topological phonon mode will be highly stable and localized at the interface. To verify the stability and localization properties of the interface phonon mode, we performed MD simulations to study the energy transfer along the C-C interface.

We created a SiC sheet of size $200 \vec{a}_1 \times 20 \vec{a}_2$ as shown in Fig.~\ref{fig_md_sic}~(a). The C-C interface is along the horizontal direction in the middle of the SiC sheet.  Waves with specified frequency were generated at the left end of the interface by driving one atom to oscillate at the given frequency for 30 cycles, which is modulated by the Hanning window. The oscillation of this atom is driven in the z-direction only, as we are only interested in the interface phonon branch crossing over the lower-frequency gap that corresponds to the z-directional vibration.

Some MD snapshots are presented in Fig.~\ref{fig_md_sic} for the energy transfer along the C-C interface of the SiC. In Fig.~\ref{fig_md_sic}~(a), the vibrational energy excited at frequency $\omega=300$~{cm$^{-1}$} (which is within the interface branch crossing over the lower-frequency gap) travels along the interface from left to right. \rev{Note that we have combined two snapshots at different time in the figure.} The velocity of the wave packet from the MD simulation is 20.5~{\AA/ps}, which is almost the same as the group velocity of 21.5~{\AA/ps} for the interface phonon with frequency $\omega=300$~{cm$^{-1}$}. In Fig.~\ref{fig_md_sic}~(b), the vibrational energy excited at frequency $\omega=100$~{cm$^{-1}$} is a normal phonon mode, which falls outside of the interface phonon branch crossing over the lower-frequency range. This normal mode is spatially extended and can travel in the space. Fig.~\ref{fig_md_sic}~(c) shows that it is rather difficult to inject energy into the SiC by vibrating at frequency $\omega=700$~{cm$^{-1}$}, which is outside the eigenfrequency range of the z-directional vibration in the SiC.

\section{Conclusion}
In conclusion, we have demonstrated the existence of topologically protected interfacial phonon modes in a monolayer, two-dimensional h-BN and SiC sheets.  The phonon dispersion of the topological interface mode crosses over the frequency gap corresponding to the in-plane vibration for h-BN, which is opened by breaking the inversion symmetry of the primitive unit cell for h-BN. In SiC, the topological interface mode exists in the frequency gap corresonding to the out-of-plane vibrations, but there is no topological phonon branch crossing over the frequency for the in-plane vibrations, because of the strong breaking of the inversion symmetry in SiC. The topological interface mode is isolated from the other phonon modes by the energy conservation law of the phonon scattering mechanism, which leads to the energy associated with the topological interface mode being highly localized at the interface both spatially and temporally, while being insensitive to defects such as sharp corners. These results demonstrate the possibilities of novel physical phenomena that may emerge in two-dimensional topological nanomechanics.  

\textbf{Acknowledgements} The work is supported by the Recruitment Program of Global Youth Experts of China, the National Natural Science Foundation of China (NSFC) under Grant No. 11504225, the start-up funding from Shanghai University, and the Innovation Program of Shanghai Municipal Education Commission under Grant No. 2017-01-07-00-09-E00019. HSP acknowledges the support of the Mechanical Engineering department at Boston University.

\textbf{Competing financial interests} The authors declare no competing financial interests.


\begin{thebibliography}{10}
\expandafter\ifx\csname url\endcsname\relax
  \def\url#1{\texttt{#1}}\fi
\expandafter\ifx\csname urlprefix\endcsname\relax\def\urlprefix{URL }\fi
\providecommand{\bibinfo}[2]{#2}
\providecommand{\eprint}[2][]{\url{#2}}

\bibitem{KaneCL2005prl}
\bibinfo{author}{Kane, C.~L.} \& \bibinfo{author}{Mele, E.~J.}
\newblock \bibinfo{title}{Quantum spin hall effect in graphene}.
\newblock \emph{\bibinfo{journal}{Phys. Rev. Lett.}}
  \textbf{\bibinfo{volume}{95}}, \bibinfo{pages}{226801}
  (\bibinfo{year}{2005}).

\bibitem{HasanMZ2010rmp}
\bibinfo{author}{Hasan, M.~Z.} \& \bibinfo{author}{Kane, C.~L.}
\newblock \bibinfo{title}{Colloquium: Topological insulators}.
\newblock \emph{\bibinfo{journal}{Rev. Mod. Phys.}}
  \textbf{\bibinfo{volume}{82}}, \bibinfo{pages}{3045} (\bibinfo{year}{2010}).

\bibitem{WangJ2017nm}
\bibinfo{author}{Wang, J.} \& \bibinfo{author}{Zhang, S.-C.}
\newblock \bibinfo{title}{Topological states of condensed matter}.
\newblock \emph{\bibinfo{journal}{Nat. Mater.}} \textbf{\bibinfo{volume}{16}},
  \bibinfo{pages}{1062} (\bibinfo{year}{2017}).

\bibitem{mooreNATURE2010}
\bibinfo{author}{Moore, J.~E.}
\newblock \bibinfo{title}{The birth of topological insulators}.
\newblock \emph{\bibinfo{journal}{Nature}} \textbf{\bibinfo{volume}{464}},
  \bibinfo{pages}{194--198} (\bibinfo{year}{2010}).

\bibitem{qiPT2010}
\bibinfo{author}{Qi, X.-L.} \& \bibinfo{author}{Zhang, S.-C.}
\newblock \bibinfo{title}{The quantum spin hall effect and topological
  insulators}.
\newblock \emph{\bibinfo{journal}{Physics Today}}
  \textbf{\bibinfo{volume}{63}}, \bibinfo{pages}{33--38}
  (\bibinfo{year}{2010}).

\bibitem{bernevigSCIENCE2006}
\bibinfo{author}{Bernevig, B.~A.}, \bibinfo{author}{Hughes, T.~L.} \&
  \bibinfo{author}{Zhang, S.-C.}
\newblock \bibinfo{title}{Quantum spin hall effect and topological phase
  transition in {H}g{T}e quantum wells}.
\newblock \emph{\bibinfo{journal}{Science}} \textbf{\bibinfo{volume}{314}},
  \bibinfo{pages}{1757--1761} (\bibinfo{year}{2006}).

\bibitem{kanePRL2005a}
\bibinfo{author}{Kane, C.~L.} \& \bibinfo{author}{Mele, E.~J.}
\newblock \bibinfo{title}{Quantum spin hall effect in graphene}.
\newblock \emph{\bibinfo{journal}{Physical Review Letters}}
  \textbf{\bibinfo{volume}{95}}, \bibinfo{pages}{226801}
  (\bibinfo{year}{2005}).

\bibitem{HaldaneFDM2008prl}
\bibinfo{author}{Haldane, F. D.~M.} \& \bibinfo{author}{Raghu, S.}
\newblock \bibinfo{title}{Possible realization of directional optical
  waveguides in photonic crystals with broken time-reversal symmetry}.
\newblock \emph{\bibinfo{journal}{Phys. Rev. Lett.}}
  \textbf{\bibinfo{volume}{100}}, \bibinfo{pages}{013904}
  (\bibinfo{year}{2008}).

\bibitem{LuL2014npho}
\bibinfo{author}{Lu, L.}, \bibinfo{author}{Joannopoulos, J.~D.} \&
  \bibinfo{author}{Soljacic, M.}
\newblock \bibinfo{title}{Topological photonics}.
\newblock \emph{\bibinfo{journal}{Nat. Photonics}}
  \textbf{\bibinfo{volume}{8}}, \bibinfo{pages}{821--829}
  (\bibinfo{year}{2014}).

\bibitem{ChenXD2017prb}
\bibinfo{author}{Chen, X.-D.}, \bibinfo{author}{Zhao, F.-L.},
  \bibinfo{author}{Chen, M.} \& \bibinfo{author}{Dong, J.-W.}
\newblock \bibinfo{title}{Valley-contrasting physics in all-dielectric photonic
  crystals: Orbital angular momentum and topological propagation}.
\newblock \emph{\bibinfo{journal}{Phys. Rev. B}} \textbf{\bibinfo{volume}{96}},
  \bibinfo{pages}{020202(R)} (\bibinfo{year}{2017}).

\bibitem{PeanoV2015prx}
\bibinfo{author}{Peano, V.}, \bibinfo{author}{Brendel, C.},
  \bibinfo{author}{Schmidt, M.} \& \bibinfo{author}{Marquardt, F.}
\newblock \bibinfo{title}{Topological phases of sound and light}.
\newblock \emph{\bibinfo{journal}{Phys. Rev. X}} \textbf{\bibinfo{volume}{5}},
  \bibinfo{pages}{031011} (\bibinfo{year}{2015}).

\bibitem{swinteckJAP2015}
\bibinfo{author}{Swinteck, N.} \emph{et~al.}
\newblock \bibinfo{title}{Bulk elastic waves with unidirectional
  backscattering-immune topological states in a time-dependent superlattice}.
\newblock \emph{\bibinfo{journal}{Journal of Applied Physics}}
  \textbf{\bibinfo{volume}{118}}, \bibinfo{pages}{063103}
  (\bibinfo{year}{2015}).

\bibitem{nassarJMPS2017}
\bibinfo{author}{Nassar, H.}, \bibinfo{author}{Xu, X.~C.},
  \bibinfo{author}{Norris, A.~N.} \& \bibinfo{author}{Huang, G.~L.}
\newblock \bibinfo{title}{Modulated phononic crystals: non-reciprocal wave
  propagation and willis crystals}.
\newblock \emph{\bibinfo{journal}{Journal of the Mechanics and Physics of
  Solids}} \textbf{\bibinfo{volume}{101}}, \bibinfo{pages}{10--29}
  (\bibinfo{year}{2017}).

\bibitem{prodanPRL2009}
\bibinfo{author}{Prodan, E.} \& \bibinfo{author}{Prodan, C.}
\newblock \bibinfo{title}{Topological phonon modes and their role in dynamic
  instability of microtubules}.
\newblock \emph{\bibinfo{journal}{Physical Review Letters}}
  \textbf{\bibinfo{volume}{103}}, \bibinfo{pages}{248101}
  (\bibinfo{year}{2009}).

\bibitem{nashPNAS2015}
\bibinfo{author}{Nash, L.~M.} \emph{et~al.}
\newblock \bibinfo{title}{Topological mechanics of gyroscopic metamaterials}.
\newblock \emph{\bibinfo{journal}{Proceedings of the National Academy of
  Science}} \textbf{\bibinfo{volume}{112}}, \bibinfo{pages}{14495--14500}
  (\bibinfo{year}{2015}).

\bibitem{wangPRL2015}
\bibinfo{author}{Wang, P.}, \bibinfo{author}{Lu, L.} \&
  \bibinfo{author}{Bertoldi, K.}
\newblock \bibinfo{title}{Topological phononic crystals with one-way elastic
  edge waves}.
\newblock \emph{\bibinfo{journal}{Physical Review Letters}}
  \textbf{\bibinfo{volume}{115}}, \bibinfo{pages}{104302}
  (\bibinfo{year}{2015}).

\bibitem{kariyadoSR2015}
\bibinfo{author}{Kariyado, T.} \& \bibinfo{author}{Hatsugai, Y.}
\newblock \bibinfo{title}{Manipulation of dirac cones in mechanical graphene}.
\newblock \emph{\bibinfo{journal}{Scientific Reports}}
  \textbf{\bibinfo{volume}{5}}, \bibinfo{pages}{18107} (\bibinfo{year}{2015}).

\bibitem{yangPRL2015}
\bibinfo{author}{Yang, Z.} \emph{et~al.}
\newblock \bibinfo{title}{Topological acoustics}.
\newblock \emph{\bibinfo{journal}{Physical Review Letters}}
  \textbf{\bibinfo{volume}{114}}, \bibinfo{pages}{114301}
  (\bibinfo{year}{2015}).

\bibitem{chenPRA2016}
\bibinfo{author}{Chen, Z.-G.} \& \bibinfo{author}{Wu, Y.}
\newblock \bibinfo{title}{Tunable topological phononic crystals}.
\newblock \emph{\bibinfo{journal}{Physical Review Applied}}
  \textbf{\bibinfo{volume}{5}}, \bibinfo{pages}{054021} (\bibinfo{year}{2016}).

\bibitem{khanikaevNC2015}
\bibinfo{author}{Khanikaev, A.~B.}, \bibinfo{author}{Fleury, R.},
  \bibinfo{author}{Mousavi, S.~H.} \& \bibinfo{author}{Alu, A.}
\newblock \bibinfo{title}{Topologically robust sound propagation in an
  angular-momentum-biased graphene-like resonator lattice}.
\newblock \emph{\bibinfo{journal}{Nature Communications}}
  \textbf{\bibinfo{volume}{6}}, \bibinfo{pages}{8260} (\bibinfo{year}{2015}).

\bibitem{mousaviNC2015}
\bibinfo{author}{Mousavi, S.~H.}, \bibinfo{author}{Khanikaev, A.~B.} \&
  \bibinfo{author}{Wang, Z.}
\newblock \bibinfo{title}{Topologically protected elastic waves in phononic
  metamaterials}.
\newblock \emph{\bibinfo{journal}{Nature Communications}}
  \textbf{\bibinfo{volume}{6}}, \bibinfo{pages}{8682} (\bibinfo{year}{2015}).

\bibitem{susstrunkPNAS2016}
\bibinfo{author}{Susstrunk, R.} \& \bibinfo{author}{Huber, S.~D.}
\newblock \bibinfo{title}{Classification of topological phonons in linear
  mechanical metamaterials}.
\newblock \emph{\bibinfo{journal}{Proceedings of the National Academy of
  Science}} \textbf{\bibinfo{volume}{113}}, \bibinfo{pages}{E4767--E4775}
  (\bibinfo{year}{2016}).

\bibitem{susstrunkSCIENCE2015}
\bibinfo{author}{Susstrunk, R.} \& \bibinfo{author}{Huber, S.~D.}
\newblock \bibinfo{title}{Observation of phononic helical edge states in a
  mechanical topological insulator}.
\newblock \emph{\bibinfo{journal}{Science}} \textbf{\bibinfo{volume}{349}},
  \bibinfo{pages}{47--50} (\bibinfo{year}{2015}).

\bibitem{palJAP2016}
\bibinfo{author}{Pal, R.~K.}, \bibinfo{author}{Schaeffer, M.} \&
  \bibinfo{author}{Ruzzene, M.}
\newblock \bibinfo{title}{Helical edge states and topological phase transitions
  in phononic systems using bi-layered lattices}.
\newblock \emph{\bibinfo{journal}{Journal of Applied Physics}}
  \textbf{\bibinfo{volume}{119}}, \bibinfo{pages}{084305}
  (\bibinfo{year}{2016}).

\bibitem{palARXIV2017}
\bibinfo{author}{Pal, R.~K.}, \bibinfo{author}{Vila, J.},
  \bibinfo{author}{Leamy, M.} \& \bibinfo{author}{Ruzzene, M.}
\newblock \bibinfo{title}{Amplitude-dependent topological edge states in
  nonlinear phononic lattices}.
\newblock \emph{\bibinfo{journal}{Arxiv}} \bibinfo{pages}{1705.01118}
  (\bibinfo{year}{2017}).

\bibitem{yuARXIV2017}
\bibinfo{author}{Yu, S.-Y.} \emph{et~al.}
\newblock \bibinfo{title}{A monolithic topologically protected phononic
  circuit}.
\newblock \emph{\bibinfo{journal}{Arxiv}} \bibinfo{pages}{1707.04901}
  (\bibinfo{year}{2017}).

\bibitem{prodanNC2017}
\bibinfo{author}{Prodan, E.}, \bibinfo{author}{Dobiszewski, K.},
  \bibinfo{author}{Kanwal, A.}, \bibinfo{author}{Palmieri, J.} \&
  \bibinfo{author}{Prodan, C.}
\newblock \bibinfo{title}{Dynamical majorana edge modes in a broad class of
  topological mechanical systems}.
\newblock \emph{\bibinfo{journal}{Nature Communications}}
  \textbf{\bibinfo{volume}{8}}, \bibinfo{pages}{14587} (\bibinfo{year}{2017}).

\bibitem{husseinAMR2014}
\bibinfo{author}{Hussein, M.~I.}, \bibinfo{author}{Leamy, M.~J.} \&
  \bibinfo{author}{Ruzzenne, M.}
\newblock \bibinfo{title}{Dynamics of phononic materials and structures:
  historical origins, recent progress, and future outlook}.
\newblock \emph{\bibinfo{journal}{Applied Mechanics Reviews}}
  \textbf{\bibinfo{volume}{66}}, \bibinfo{pages}{040802}
  (\bibinfo{year}{2014}).

\bibitem{heNP2016}
\bibinfo{author}{He, C.} \emph{et~al.}
\newblock \bibinfo{title}{Acoustic topological insulator and robust one-way
  sound transport}.
\newblock \emph{\bibinfo{journal}{Nature Physics}}
  \textbf{\bibinfo{volume}{12}}, \bibinfo{pages}{1124--1129}
  (\bibinfo{year}{2016}).

\bibitem{cummerNRM2016}
\bibinfo{author}{Cummer, S.~A.}, \bibinfo{author}{Christensen, J.} \&
  \bibinfo{author}{Alu, A.}
\newblock \bibinfo{title}{Controlling sound with acoustic metamaterials}.
\newblock \emph{\bibinfo{journal}{Nature Reviews Materials}}
  \textbf{\bibinfo{volume}{1}}, \bibinfo{pages}{1--13} (\bibinfo{year}{2016}).

\bibitem{palNJP2017}
\bibinfo{author}{Pal, R.~K.} \& \bibinfo{author}{Ruzzene, M.}
\newblock \bibinfo{title}{Edge waves in plates with resonators: an elastic
  analogue of the quantum valley hall effect}.
\newblock \emph{\bibinfo{journal}{New Journal of Physics}}
  \textbf{\bibinfo{volume}{19}}, \bibinfo{pages}{025001}
  (\bibinfo{year}{2017}).

\bibitem{xiaoNP2015}
\bibinfo{author}{Xiao, M.} \emph{et~al.}
\newblock \bibinfo{title}{Geometric phase and band inversion in periodic
  acoustic systems}.
\newblock \emph{\bibinfo{journal}{Nature Physics}}
  \textbf{\bibinfo{volume}{11}}, \bibinfo{pages}{240--244}
  (\bibinfo{year}{2015}).

\bibitem{renRPP2016}
\bibinfo{author}{Ren, Y.}, \bibinfo{author}{Qiao, Z.} \& \bibinfo{author}{Niu,
  Q.}
\newblock \bibinfo{title}{Topological phases in two-dimensional materials: a
  review}.
\newblock \emph{\bibinfo{journal}{Reports on Progress in Physics}}
  \textbf{\bibinfo{volume}{79}}, \bibinfo{pages}{066501}
  (\bibinfo{year}{2016}).

\bibitem{liuARXIV2017}
\bibinfo{author}{Liu, T.-W.} \& \bibinfo{author}{Semperlotti, F.}
\newblock \bibinfo{title}{Acoustic valley-hall edge states in phononic elastic
  waveguides}.
\newblock \emph{\bibinfo{journal}{Arxiv}} \bibinfo{pages}{1708.02987}
  (\bibinfo{year}{2017}).

\bibitem{wuARXIV2017}
\bibinfo{author}{Wu, Y.}, \bibinfo{author}{Chaunsali, R.},
  \bibinfo{author}{Yasuda, H.}, \bibinfo{author}{Yu, K.} \&
  \bibinfo{author}{Yang, J.}
\newblock \bibinfo{title}{Dial-in topological metamaterials based on bistable
  stewart platform}.
\newblock \emph{\bibinfo{journal}{Arxiv}} \bibinfo{pages}{1710.00065}
  (\bibinfo{year}{2017}).

\bibitem{KosevichAM2004ltp}
\bibinfo{author}{Kosevich, A.~M.}
\newblock \bibinfo{title}{Topology and solid-state physics}.
\newblock \emph{\bibinfo{journal}{Low Temp. Phys.}}
  \textbf{\bibinfo{volume}{30}}, \bibinfo{pages}{97} (\bibinfo{year}{2004}).

\bibitem{ZhangL2010prl}
\bibinfo{author}{Zhang, L.}, \bibinfo{author}{Ren, J.}, \bibinfo{author}{Wang,
  J.-S.} \& \bibinfo{author}{Li, B.}
\newblock \bibinfo{title}{Topological nature of the phonon hall effect}.
\newblock \emph{\bibinfo{journal}{Phys. Rev. Lett.}}
  \textbf{\bibinfo{volume}{105}}, \bibinfo{pages}{225901}
  (\bibinfo{year}{2010}).

\bibitem{SusstrunkR2016pnas}
\bibinfo{author}{Susstrunk, R.} \& \bibinfo{author}{Huber, S.~D.}
\newblock \bibinfo{title}{Classification of topological phonons in linear
  mechanical metamaterials}.
\newblock \emph{\bibinfo{journal}{Proc. Natl. Acad. Sci.}}
  \textbf{\bibinfo{volume}{113}}, \bibinfo{pages}{E4767--E4775}
  (\bibinfo{year}{2016}).

\bibitem{LiuY2017prb}
\bibinfo{author}{Liu, Y.}, \bibinfo{author}{Xu, Y.}, \bibinfo{author}{Zhang,
  S.-C.} \& \bibinfo{author}{Duan, W.}
\newblock \bibinfo{title}{Model for topological phononics and phonon diode}.
\newblock \emph{\bibinfo{journal}{Phys. Rev. B}} \textbf{\bibinfo{volume}{96}},
  \bibinfo{pages}{064106} (\bibinfo{year}{2017}).

\bibitem{YangZ2015prl}
\bibinfo{author}{Yang, Z.} \emph{et~al.}
\newblock \bibinfo{title}{Topological acoustics}.
\newblock \emph{\bibinfo{journal}{Phys. Rev. Lett.}}
  \textbf{\bibinfo{volume}{114}}, \bibinfo{pages}{114301}
  (\bibinfo{year}{2015}).

\bibitem{PalRK2016jap}
\bibinfo{author}{Pal, R.~K.}, \bibinfo{author}{Schaeffer, M.} \&
  \bibinfo{author}{Ruzzene, M.}
\newblock \bibinfo{title}{Helical edge states and topological phase transitions
  in phononic systems using bi-layered lattices}.
\newblock \emph{\bibinfo{journal}{J. Appl. Phys.}}
  \textbf{\bibinfo{volume}{119}}, \bibinfo{pages}{084305}
  (\bibinfo{year}{2016}).

\bibitem{SussmanDM2016sm}
\bibinfo{author}{Sussman, D.~M.}, \bibinfo{author}{Stenull, O.} \&
  \bibinfo{author}{Lubensky, T.~C.}
\newblock \bibinfo{title}{Topological boundary modes in jammed matter}.
\newblock \emph{\bibinfo{journal}{Soft Matter}} \textbf{\bibinfo{volume}{12}},
  \bibinfo{pages}{6079--6087} (\bibinfo{year}{2016}).

\bibitem{JiWC2017cpl}
\bibinfo{author}{Ji, W.-C.} \& \bibinfo{author}{Shi, J.-R.}
\newblock \bibinfo{title}{Topological phonon modes in a two-dimensional wigner
  crystal}.
\newblock \emph{\bibinfo{journal}{Chin. Phys. Lett.}}
  \textbf{\bibinfo{volume}{34}}, \bibinfo{pages}{036301}
  (\bibinfo{year}{2017}).

\bibitem{SunK2012pnas}
\bibinfo{author}{Sun, K.}, \bibinfo{author}{Souslov, A.}, \bibinfo{author}{Mao,
  X.} \& \bibinfo{author}{Lubensky, T.~C.}
\newblock \bibinfo{title}{Surface phonons, elastic response, and conformal
  invariance in twisted kagome lattices}.
\newblock \emph{\bibinfo{journal}{Proc. Natl. Acad. Sci.}}
  \textbf{\bibinfo{volume}{109}}, \bibinfo{pages}{12369--12374}
  (\bibinfo{year}{2012}).

\bibitem{KaneCL2014nphy}
\bibinfo{author}{Kane, C.~L.} \& \bibinfo{author}{Lubensky, T.~C.}
\newblock \bibinfo{title}{Topological boundary modes in isostatic lattices}.
\newblock \emph{\bibinfo{journal}{Nat. Phys.}} \textbf{\bibinfo{volume}{10}},
  \bibinfo{pages}{39--45} (\bibinfo{year}{2014}).

\bibitem{RocklinDZ2017nc}
\bibinfo{author}{Rocklin, D.~Z.}, \bibinfo{author}{Zhou, S.},
  \bibinfo{author}{Sun, K.} \& \bibinfo{author}{Mao, X.}
\newblock \bibinfo{title}{Transformable topological mechanical metamaterials}.
\newblock \emph{\bibinfo{journal}{Nat. Commun.}} \textbf{\bibinfo{volume}{8}},
  \bibinfo{pages}{14201} (\bibinfo{year}{2017}).

\bibitem{ProdanE2009prl}
\bibinfo{author}{Prodan, E.} \& \bibinfo{author}{Prodan, C.}
\newblock \bibinfo{title}{Topological phonon modes and their role in dynamic
  instability of microtubules}.
\newblock \emph{\bibinfo{journal}{Phys. Rev. Lett.}}
  \textbf{\bibinfo{volume}{103}}, \bibinfo{pages}{248101}
  (\bibinfo{year}{2009}).

\bibitem{BergN2011pre}
\bibinfo{author}{Berg, N.}, \bibinfo{author}{Joel, K.},
  \bibinfo{author}{Koolyk, M.} \& \bibinfo{author}{Prodan, E.}
\newblock \bibinfo{title}{Topological phonon modes in filamentary structures}.
\newblock \emph{\bibinfo{journal}{Phys. Rev. E}} \textbf{\bibinfo{volume}{83}},
  \bibinfo{pages}{021913} (\bibinfo{year}{2011}).

\bibitem{ProdanE2017nc}
\bibinfo{author}{Prodan, E.}, \bibinfo{author}{Dobiszewski, K.},
  \bibinfo{author}{Kanwal, A.}, \bibinfo{author}{Palmieri, J.} \&
  \bibinfo{author}{Prodan, C.}
\newblock \bibinfo{title}{Dynamical majorana edge modes in a broad class of
  topological mechanical systems}.
\newblock \emph{\bibinfo{journal}{Nat. Commun.}} \textbf{\bibinfo{volume}{8}},
  \bibinfo{pages}{14587} (\bibinfo{year}{2017}).

\bibitem{PauloseJ2015nphy}
\bibinfo{author}{Paulose, J.}, \bibinfo{author}{ge~Chen, B.~G.} \&
  \bibinfo{author}{Vitelli, V.}
\newblock \bibinfo{title}{Topological modes bound to dislocations in mechanical
  metamaterials}.
\newblock \emph{\bibinfo{journal}{Nat. Phys.}} \textbf{\bibinfo{volume}{11}},
  \bibinfo{pages}{153--156} (\bibinfo{year}{2015}).

\bibitem{LubenskyTC2015rpp}
\bibinfo{author}{Lubensky, T.~C.}, \bibinfo{author}{Kane, C.~L.},
  \bibinfo{author}{Mao, X.}, \bibinfo{author}{Souslov, A.} \&
  \bibinfo{author}{Sun, K.}
\newblock \bibinfo{title}{Phonons and elasticity in critically coordinated
  lattices}.
\newblock \emph{\bibinfo{journal}{Rep. Prog. Phys.}}
  \textbf{\bibinfo{volume}{78}}, \bibinfo{pages}{073901}
  (\bibinfo{year}{2015}).

\bibitem{PauloseJ2015pnas}
\bibinfo{author}{Paulose, J.}, \bibinfo{author}{Meeussen, A.~S.} \&
  \bibinfo{author}{Vitelli, V.}
\newblock \bibinfo{title}{Selective buckling via states of self-stress in
  topological metamaterials}.
\newblock \emph{\bibinfo{journal}{Proc. Natl. Acad. Sci.}}
  \textbf{\bibinfo{volume}{112}}, \bibinfo{pages}{7639--7644}
  (\bibinfo{year}{2015}).

\bibitem{RocklinDZ2016prl}
\bibinfo{author}{Rocklin, D.~Z.}, \bibinfo{author}{Chen, B.~G.},
  \bibinfo{author}{Falk, M.}, \bibinfo{author}{Vitelli, V.} \&
  \bibinfo{author}{Lubensky, T.~C.}
\newblock \bibinfo{title}{Mechanical weyl modes in topological maxwell
  lattices}.
\newblock \emph{\bibinfo{journal}{Phys. Rev. Lett.}}
  \textbf{\bibinfo{volume}{116}}, \bibinfo{pages}{135503}
  (\bibinfo{year}{2016}).

\bibitem{StenullO2016prl}
\bibinfo{author}{Stenull, O.}, \bibinfo{author}{Kane, C.~L.} \&
  \bibinfo{author}{Lubensky, T.~C.}
\newblock \bibinfo{title}{Topological phonons and weyl lines in three
  dimensions}.
\newblock \emph{\bibinfo{journal}{Phys. Rev. Lett.}}
  \textbf{\bibinfo{volume}{117}}, \bibinfo{pages}{068001}
  (\bibinfo{year}{2016}).

\bibitem{PoHC2016prb}
\bibinfo{author}{Po, H.~C.}, \bibinfo{author}{Bahri, Y.} \&
  \bibinfo{author}{Vishwanath, A.}
\newblock \bibinfo{title}{Phonon analog of topological nodal semimetals}.
\newblock \emph{\bibinfo{journal}{Phys. Rev. B}} \textbf{\bibinfo{volume}{93}},
  \bibinfo{pages}{205158} (\bibinfo{year}{2016}).

\bibitem{XiaoD2007prl}
\bibinfo{author}{Xiao, D.}, \bibinfo{author}{Yao, W.} \& \bibinfo{author}{Niu,
  Q.}
\newblock \bibinfo{title}{Valley-contrasting physics in graphene: Magnetic
  moment and topological transport}.
\newblock \emph{\bibinfo{journal}{Phys. Rev. Lett.}}
  \textbf{\bibinfo{volume}{99}}, \bibinfo{pages}{236809}
  (\bibinfo{year}{2007}).

\bibitem{ZhangF2011prl}
\bibinfo{author}{Zhang, F.}, \bibinfo{author}{Jung, J.},
  \bibinfo{author}{Fiete, G.~A.}, \bibinfo{author}{Niu, Q.} \&
  \bibinfo{author}{MacDonald, A.~H.}
\newblock \bibinfo{title}{Spontaneous quantum hall states in chirally stacked
  few-layer graphene systems}.
\newblock \emph{\bibinfo{journal}{Phys. Rev. Lett.}}
  \textbf{\bibinfo{volume}{106}}, \bibinfo{pages}{156801}
  (\bibinfo{year}{2011}).

\bibitem{ZhangF2013pnas}
\bibinfo{author}{Zhang, F.}, \bibinfo{author}{MacDonald, A.~H.} \&
  \bibinfo{author}{Mele, E.~J.}
\newblock \bibinfo{title}{Valley chern numbers and boundary modes in gapped
  bilayer graphene}.
\newblock \emph{\bibinfo{journal}{Proc. Natl. Acad. Sci.}}
  \textbf{\bibinfo{volume}{110}}, \bibinfo{pages}{10546--10551}
  (\bibinfo{year}{2013}).

\bibitem{brennerJPCM2002}
\bibinfo{author}{Brenner, D.~W.} \emph{et~al.}
\newblock \bibinfo{title}{A second-generation reactive empirical bond order
  ({REBO}) potential energy expression for hydrocarbons}.
\newblock \emph{\bibinfo{journal}{J. Phys.: Condens. Matter}}
  \textbf{\bibinfo{volume}{14}}, \bibinfo{pages}{783--802}
  (\bibinfo{year}{2002}).

\bibitem{LindsayL2011prb}
\bibinfo{author}{Lindsay, L.} \& \bibinfo{author}{Broido, D.~A.}
\newblock \bibinfo{title}{Enhanced thermal conductivity and isotope effect in
  single-layer hexagonal boron nitride}.
\newblock \emph{\bibinfo{journal}{Phys. Rev. B}} \textbf{\bibinfo{volume}{84}},
  \bibinfo{pages}{155421} (\bibinfo{year}{2011}).

\bibitem{TersoffJ5}
\bibinfo{author}{Tersoff, J.}
\newblock \bibinfo{title}{Modeling solid-state chemistry: Interatomic
  potentials for multicomponent systems}.
\newblock \emph{\bibinfo{journal}{Phys. Rev. B}} \textbf{\bibinfo{volume}{39}},
  \bibinfo{pages}{5566--5568} (\bibinfo{year}{1989}).

\bibitem{gulp}
\bibinfo{author}{Gale, J.~D.}
\newblock \bibinfo{title}{Gulp: A computer program for the symmetry-adapted
  simulation of solids}.
\newblock \emph{\bibinfo{journal}{J. Chem. Soc., Faraday Trans.}}
  \textbf{\bibinfo{volume}{93}}, \bibinfo{pages}{629--637. Code available from
  https://projects.ivec.org/gulp/} (\bibinfo{year}{1997}).

\bibitem{PlimptonSJ}
\bibinfo{author}{Plimpton, S.~J.}
\newblock \bibinfo{title}{Fast parallel algorithms for short-range molecular
  dynamics}.
\newblock \emph{\bibinfo{journal}{J. Comput. Phys.}}
  \textbf{\bibinfo{volume}{117}}, \bibinfo{pages}{1--19}
  (\bibinfo{year}{1995}).

\bibitem{ovito}
\bibinfo{author}{Stukowski, A.}
\newblock \bibinfo{title}{Visualization and analysis of atomistic simulation
  data with ovito - the open visualization tool}.
\newblock \emph{\bibinfo{journal}{Modelling Simul. Mater. Sci. Eng.}}
  \textbf{\bibinfo{volume}{18}}, \bibinfo{pages}{015012}
  (\bibinfo{year}{2010}).

\bibitem{PalRK2017njp}
\bibinfo{author}{Pal, R.~K.} \& \bibinfo{author}{Ruzzene, M.}
\newblock \bibinfo{title}{Edge waves in plates with resonators: an elastic
  analogue of the quantum valley hall effect}.
\newblock \emph{\bibinfo{journal}{New J. Phys.}} \textbf{\bibinfo{volume}{19}},
  \bibinfo{pages}{025001} (\bibinfo{year}{2017}).

\bibitem{rycerzNP2007}
\bibinfo{author}{Rycerz, A.}, \bibinfo{author}{Tworzydlo, J.} \&
  \bibinfo{author}{Beenakker, C. W.~J.}
\newblock \bibinfo{title}{Valley filter and valley valve in graphene}.
\newblock \emph{\bibinfo{journal}{Nature Physics}}
  \textbf{\bibinfo{volume}{3}}, \bibinfo{pages}{172--175}
  (\bibinfo{year}{2007}).

\bibitem{ZhuH2017arxiv}
\bibinfo{author}{Zhu, H.}, \bibinfo{author}{Liu, T.-W.} \&
  \bibinfo{author}{Semperlotti, F.}
\newblock \bibinfo{title}{Design and experimental observation of valley-hall
  edge states in diatomic-graphene-like elastic waveguides}.
\newblock \emph{\bibinfo{journal}{Preprint at
  http://arxiv.org/abs/1712.10271v1}}  (\bibinfo{year}{2017}).

\bibitem{LiuY2012acsn}
\bibinfo{author}{Liu, Y.}, \bibinfo{author}{Zou, X.} \&
  \bibinfo{author}{Yakobson, B.~I.}
\newblock \bibinfo{title}{Dislocations and grain boundaries in two-dimensional
  boron nitride}.
\newblock \emph{\bibinfo{journal}{ACS Nano}} \textbf{\bibinfo{volume}{6}},
  \bibinfo{pages}{7053--7058} (\bibinfo{year}{2012}).

\bibitem{SunL2008jchemp}
\bibinfo{author}{Sun, L.} \emph{et~al.}
\newblock \bibinfo{title}{Electronic structures of sic nanoribbons}.
\newblock \emph{\bibinfo{journal}{Journal of Chemical Physics}}
  \textbf{\bibinfo{volume}{129}}, \bibinfo{pages}{174114}
  (\bibinfo{year}{2008}).

\bibitem{ShiZ2015acsn}
\bibinfo{author}{Shi, Z.}, \bibinfo{author}{Zhang, Z.},
  \bibinfo{author}{Kutana, A.} \& \bibinfo{author}{Yakobson, B.~I.}
\newblock \bibinfo{title}{Predicting two-dimensional silicon carbide
  monolayers}.
\newblock \emph{\bibinfo{journal}{ACS Nano}} \textbf{\bibinfo{volume}{9}},
  \bibinfo{pages}{9802--9809} (\bibinfo{year}{2015}).

\end{thebibliography}

\end{document}